\documentclass[twocolumn,showpacs,aps,prd,superscriptaddress,floatfix]{revtex4}
\usepackage{graphicx}
\usepackage{verbatim}
\usepackage{dcolumn}
\usepackage{multirow}
\usepackage{bm}

\newcommand{\BaBarYear}       {08}
\newcommand{\BaBarNumber}     {007}
\newcommand{\SLACPubNumber} {12516}

\newcommand{\BaBarType}      {PUB}  

\input{babarsym.tex}

\def\bcount    {\ensuremath {383 \times 10^{6}} }

\def\offlumi   {\ensuremath { 36.3 \invfb\  }}

\def\btosgamma {\ensuremath {b\to s\gamma} }
\def\bbartosbargamma {\ensuremath {\bar{b}\to \bar{s}\gamma} }
\def\ACP{\ensuremath{A_{CP}}}
\def\Xs{\ensuremath{X_s}}

\long\def\inst#1{\par\nobreak\kern 4pt\nobreak
    {\it #1}\par\vskip 10pt plus 3pt minus 3pt}

\begin{document}

{\pagestyle{empty}

\begin{flushleft}
\babar-\BaBarType-\BaBarYear/\BaBarNumber \\
SLAC-PUB-\SLACPubNumber
\end{flushleft}

\title{
  {\bf \boldmath
A Measurement of \CP\ Asymmetry in $b\to s\gamma$ using a Sum of Exclusive Final States}
}

%
\author{B.~Aubert}
\author{M.~Bona}
\author{Y.~Karyotakis}
\author{J.~P.~Lees}
\author{V.~Poireau}
\author{X.~Prudent}
\author{V.~Tisserand}
\author{A.~Zghiche}
\affiliation{Laboratoire de Physique des Particules, IN2P3/CNRS et Universit\'e de Savoie, F-74941 Annecy-Le-Vieux, France }
\author{J.~Garra~Tico}
\author{E.~Grauges}
\affiliation{Universitat de Barcelona, Facultat de Fisica, Departament ECM, E-08028 Barcelona, Spain }
% ...
\collaboration{The \babar\ Collaboration}
\noaffiliation

\date{\today}

\begin{abstract}
\noindent
We perform a measurement of the \CP\ asymmetry in $\btosgamma$ decays
using a sample of \bcount\ \BB\ events collected by the \babar\ detector
at the PEP-II asymmetric \B\ factory. We reconstruct sixteen
flavor-specific \B\ decay modes containing a high-energy
photon and a hadronic system \Xs\ containing an $s$ quark.
We measure the \CP\ asymmetry to be $-0.011\pm 0.030\stat\pm0.014\syst$ for a 
hadronic system mass between 0.6 and 2.8 \gevcc.
\end{abstract}

\pacs{13.20.-v, 13.25.Hw}

\vfill

\maketitle
}

The decay \btosgamma\ is a flavor-changing neutral current process described
by a radiative penguin diagram in the Standard Model (SM).
It is sensitive to new physics which can appear in branching
fraction or \CP\ asymmetry measurements. 
Measurements of the
branching fraction \cite{Aubert:2005cu,Aubert:2006gg}
are in good agreement with the SM \citep{Misiak:2006zs} predictions.

A \CP\ asymmetry between \btosgamma\ and \bbartosbargamma\
decays is predicted by the SM to be $\leq 1\%$~\cite{Kagan:1998bh}
but could be enhanced up to 15\% ~\cite{Wolfenstein:1994jw,Asatrian:1996as,Ciuchini:1996vw}
in models of physics beyond the SM. Existing   
measurements are consistent with zero \CP\ asymmetry with a precision of 5\% \cite{Aubert:2004hq, Nishida:2003yw}.
The increased precision obtained 
in this work allows us to better discriminate
between various theoretical models \cite{Hurth:2003dk}.

We use a sample of \bcount\ \BB\ pairs collected at the \FourS\ resonance by the 
 \babar\ detector \citep{babar} at the PEP-II $e^+e^-$  \B\ factory. 
 In addition,  we use 
\offlumi\  
collected 40 \mev\ below the \FourS\ resonance  to study backgrounds from non-\B\ decays.


We reconstruct 16 exclusive \btosgamma\ final states:
\begin{eqnarray*}
\Bm&\to&\KS\pim\gamma,\Km\piz\gamma,\Km\pip\pim\gamma,\KS\pim\piz\gamma,\\
   &   &\Km\piz\piz\gamma,\KS\pip\pim\pim\gamma,\Km\pip\pim\piz\gamma,\\
   &   &\KS\pim\piz\piz\gamma,\Km\eta\gamma,\Kp\Km\Km\gamma, \\
\Bzb&\to&\Km\pip\gamma,\Km\pip\piz\gamma,\Km\pip\pim\pip\gamma,\Km\pip\piz\piz\gamma,\\
   &   &\Km\pip\eta\gamma,\Kp\Km\Km\pip\gamma,
\end{eqnarray*}
and measure the yield asymmetry with respect to their
charge conjugate decays \bbartosbargamma.
These  modes are selected because the particles in the final state identify
                     the flavor of the \B\ meson
and they can be reconstructed with  high statistical significance.

The high-energy photon from the \B\ decay
is reconstructed from an isolated energy cluster in the
calorimeter, with a shape consistent with the electromagnetic
shower produced by a single photon, and an energy $E_\gamma^* > 1.6$ \gev\
in the \FourS\ center-of-mass (CM) frame.

The hadronic system $\Xs$, formed from the kaons and pions, is required
to have an invariant mass $M_{\Xs}$ between 0.6 and
2.8 \gevcc , corresponding to a photon energy threshold
$E_{\gamma} > $ 1.9 \gev\ in the \B\ meson rest frame.

Charged kaons are identified by combining information from the
Cherenkov detector 
and the energy-loss measurements from the tracking system.
The remaining tracks are assumed to be charged pions.
The \KS\ candidates are reconstructed by combining two oppositely charged
pions
with an invariant mass within
9 \mevcc\ of the nominal \KS\ mass \cite{Yao:2006px} and a minimum flight distance
of $2\mm$ from the primary event vertex.
Both charged and neutral kaons are required to have laboratory
momenta $\ge 0.8 \gevc$.

Neutral pions and $\eta$ candidates  are reconstructed from pairs of photons
with energies above 50 \mev in the laboratory frame and a lateral
moment \cite{Drescher:1984rt} less than 0.8.
The lateral moment  measures the spread of a shower in
the calorimeter and provides good separation between electromagnetic and hadronic showers. 
The invariant mass of the pair of photons is required to be between 115 and 150 \mevcc\  
for \piz\  candidates and between 470 and 620 \mevcc\ for  $\eta$ candidates



Monte Carlo (MC) samples based on EvtGen \cite{Lange:11} and
GEANT4 \cite{Geant4} are used to simulate the signal
and background processes and the detector response. The \btosgamma\
signal sample is generated with a photon spectrum derived
from Ref.\cite{Kagan:1998bh} assuming $m_b=4.65\gevcc$. 
The fragmentation of the \Xs\ system  is modeled using 
JETSET\cite{Sjostrand:1995iq} corrected to fit the  \babar\ data  as described later.

The background to the \B\ reconstruction is dominated by continuum
processes ($\epem\to q\bar{q}$, with $q=u,d,s,c$) that produce
a high-energy photon either by initial-state radiation or from the
decay of \piz\ and $\eta$ mesons. Continuum events tend to be less
isotropic than $B$-decay events since they
result from hadronic fragmentation of high-momentum
quarks back-to-back in the  CM frame. High-energy photons
in these events tend to be collinear with the thrust axis
formed from the rest of the event (ROE), defined as those
particles not used in reconstructing the signal \B\ candidate.
We reject such backgrounds by requiring that the cosine of
the angle between the photon and the thrust axis of the
ROE (in the CM frame) be less than 0.85.
We further reject the continuum events by requiring the ratio of
the second ($L_2$) and zeroth ($L_0$) Legendre moments
for the ROE particles with respect to the \B\ flight direction
to be smaller than 0.46. 

Continuum events with high-energy photons from \piz\ and $\eta$
decays are major backgrounds. To veto these events,
we associate each high-energy photon candidate $\gamma$
with another photon candidate $\gamma'$ in the event.
For multiple $\gamma'$ candidate in an event, we choose
the $\gamma\gamma'$ pairs whose invariant mass,
determined from adding the four vectors, is closest
to the nominal \piz\ mass (or $\eta$ mass in case of $\eta$ veto).
Events are rejected if the photon pairs are consistent with
\piz\ or $\eta$ decays based on the output of a boosted decision
tree (BDT) \cite{BDT:Freund} constructed from the energy
of the less energetic photon $\gamma'$ and $m_{\gamma\gamma'}$.

We reject the remaining continuum events
by constructing an additional BDT that combines information from a number of
variables related to the event shape,
the kinematic properties of the $B$ meson,
and the flavor-tagging~\cite{Aubert:2002ic} properties of the other $B$ meson in the event.
Examples of these variables are
the Fox-Wolfram moments \cite{Fox:1978vu}, 
and the cosine of the \B\ flight direction computed in the CM
frame with respect to the beam axis. Optimization of the selection criteria
of the \piz\ veto, $\eta$ veto, and event selection BDTs is performed
using an iterative method which maximizes the statistical signal
significance.
After the final event selection,  we reject
97\% of the continuum background while retaining 55\% of the signal events.

Fully reconstructed \btosgamma\ decays are characterized by two
kinematic variables: the beam-energy substituted mass
$\mes = \sqrt{s/4-{p_B^*}^{2}}$, and the
energy difference between the \B\ candidate and the beam energy
$\Delta E=E_B^*-\sqrt{s}/2 $, where $E_B^*$ and $p_B^*$
are the energy and momentum of the \B\ candidate in the
\epem\ CM frame, and $\sqrt{s}$ is the total CM frame energy.
Signal events are expected to have a $\Delta E$ distribution centered
near zero and a \mes\ distribution centered at the mass of the \B\ meson.
For events with multiple \B\ candidates, we select the one with the smallest $|\Delta E|$.

We perform a one-dimensional fit of \mes\ to the data in 
the entire $M_{\Xs}$ region ([0.6, 2.8] \gevcc) 
as well as in five different regions of
$M_{\Xs}$ ([0.6, 1.1], [1.1, 1.5], [1.5, 2.0] and [2.0, 2.8] \gevcc) 
 to study
whether the asymmetry has significant mass dependence.
Only candidates in the range $|\Delta E| < 0.10 \gev$
and $5.22 <\mes <5.29 \gevcc$ are considered.
Probability density functions (PDFs) are
constructed for both signal and background
in the five $M_{\Xs}$ regions. We use the
charge of the reconstructed final state (\Bm/\Bp) or the charge of
the kaon (\Bzb/\Bz) to define two flavor categories, and perform a simultaneous
fit for the flavor asymmetry in each $M_{\Xs}$ region.

The signal events are described by a function
$f(\mes)=\exp[-(\mes-\mu_0)^2/(2\sigma_{L,R}^2+\alpha_{L,R}(\mes-\mu_0)^2)]$
where the parameters are determined by an unbinned
fit to the signal MC. In the above function,
$\mu_0$ is the peak position of the distribution, $\sigma_{L,R}$ are the
widths on the left and right of the peak, and $\alpha_{L,R}$ parameterize
the tail on the left and right of the peak, respectively.

\begin{figure}[t]
\begin{center}
  \includegraphics[width=0.46\linewidth,keepaspectratio]{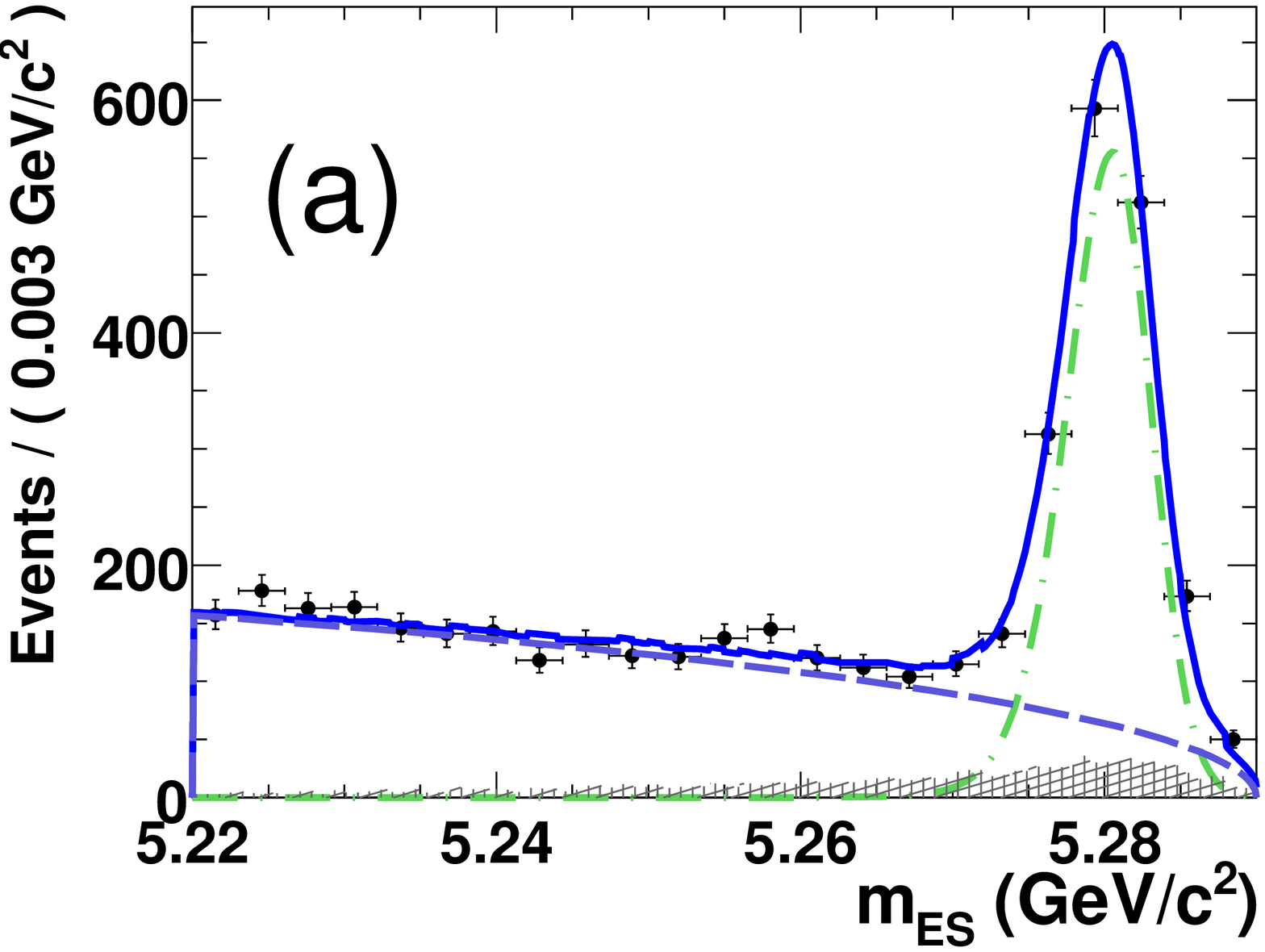}
  \includegraphics[width=0.46\linewidth,keepaspectratio]{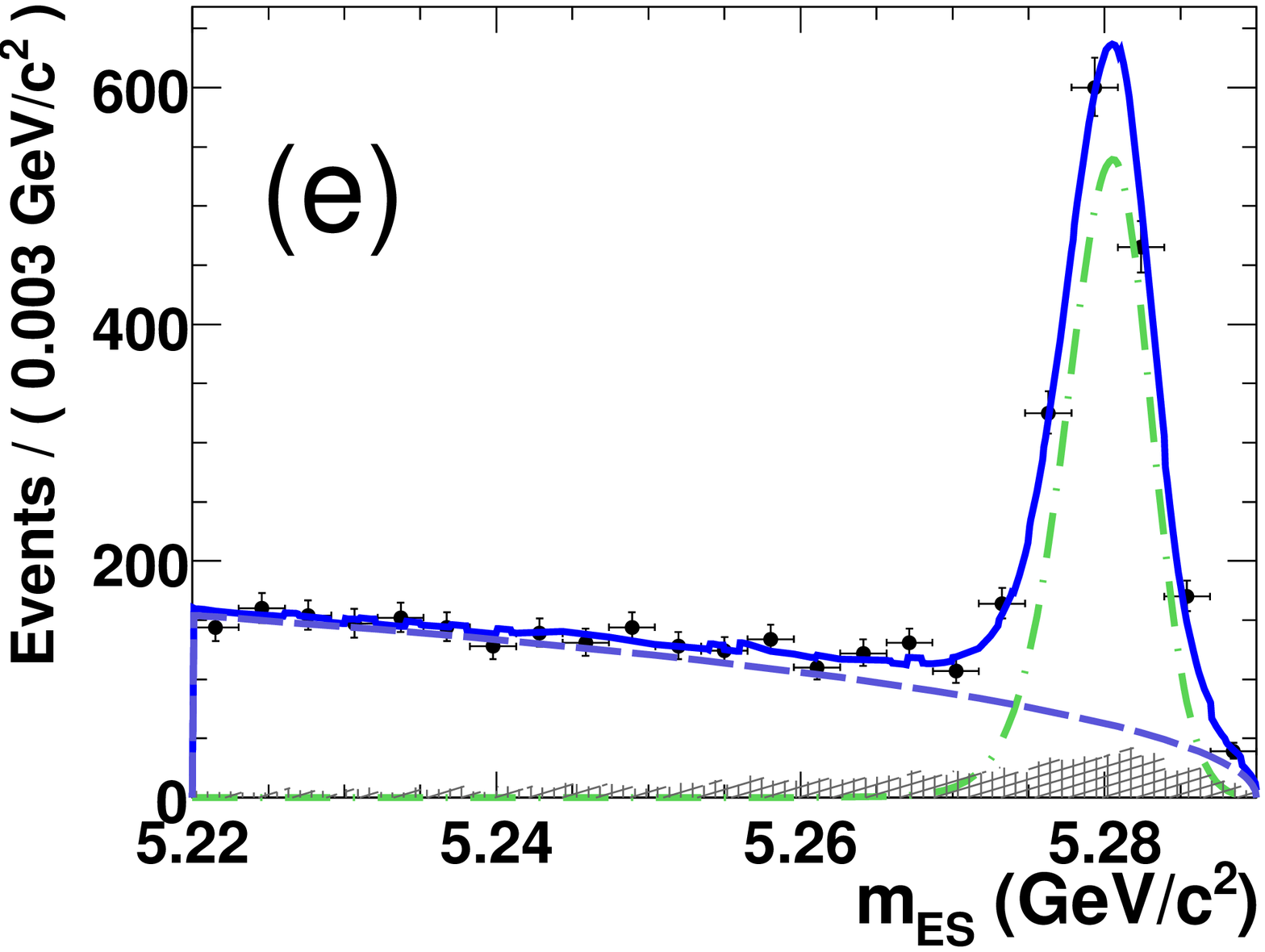}

  \includegraphics[width=0.46\linewidth,keepaspectratio]{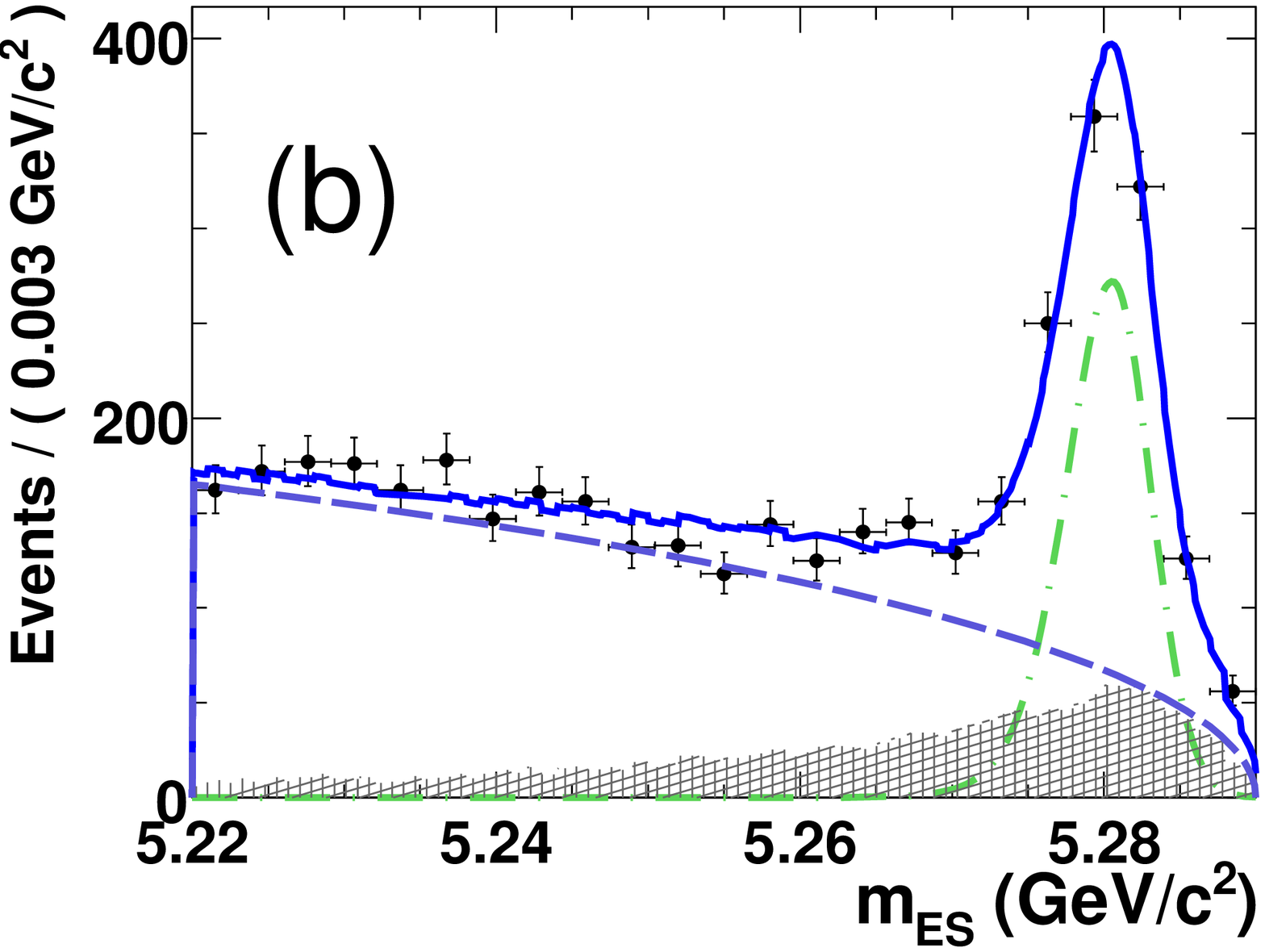}
  \includegraphics[width=0.46\linewidth,keepaspectratio]{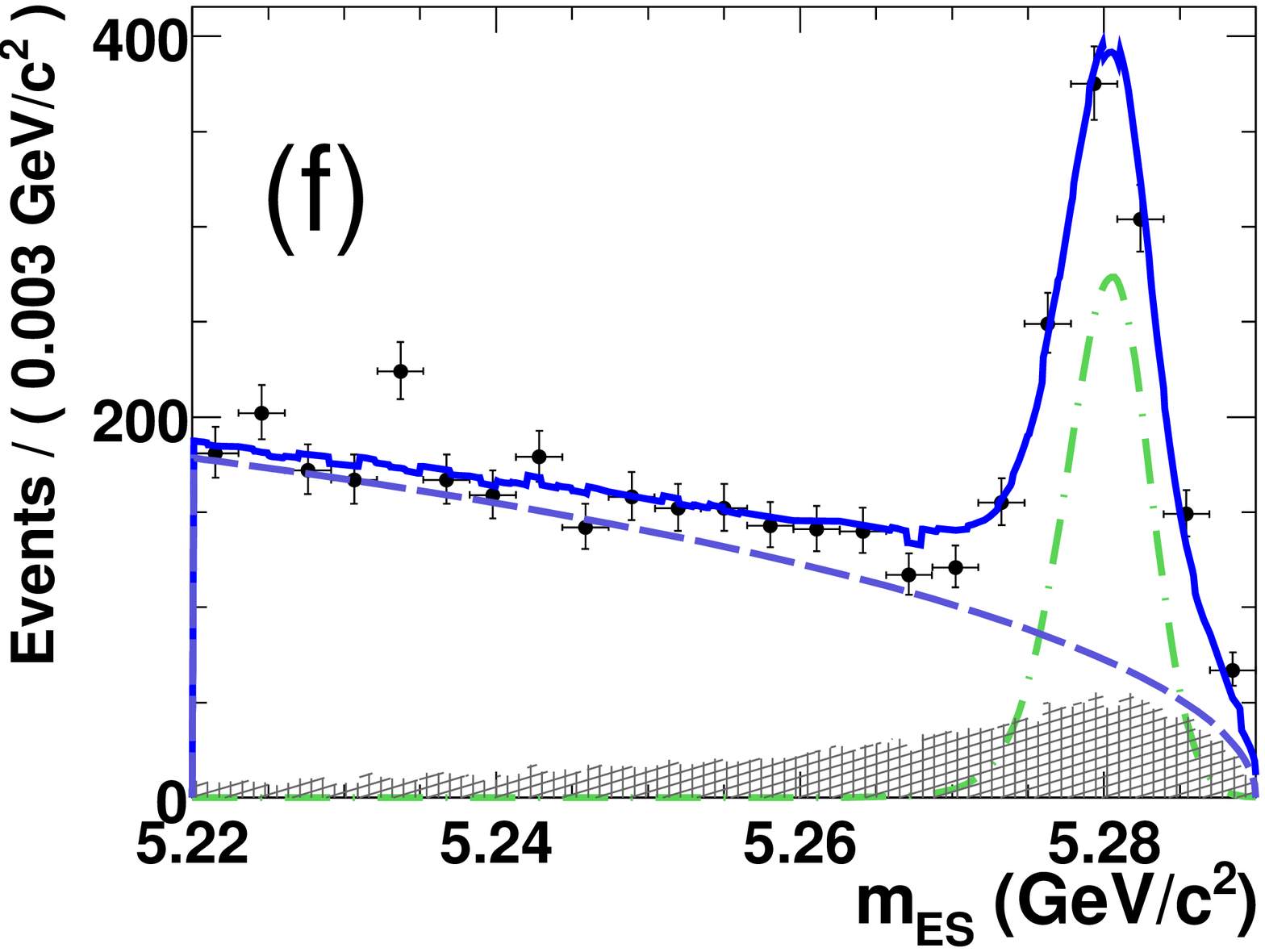}

  \includegraphics[width=0.46\linewidth,keepaspectratio]{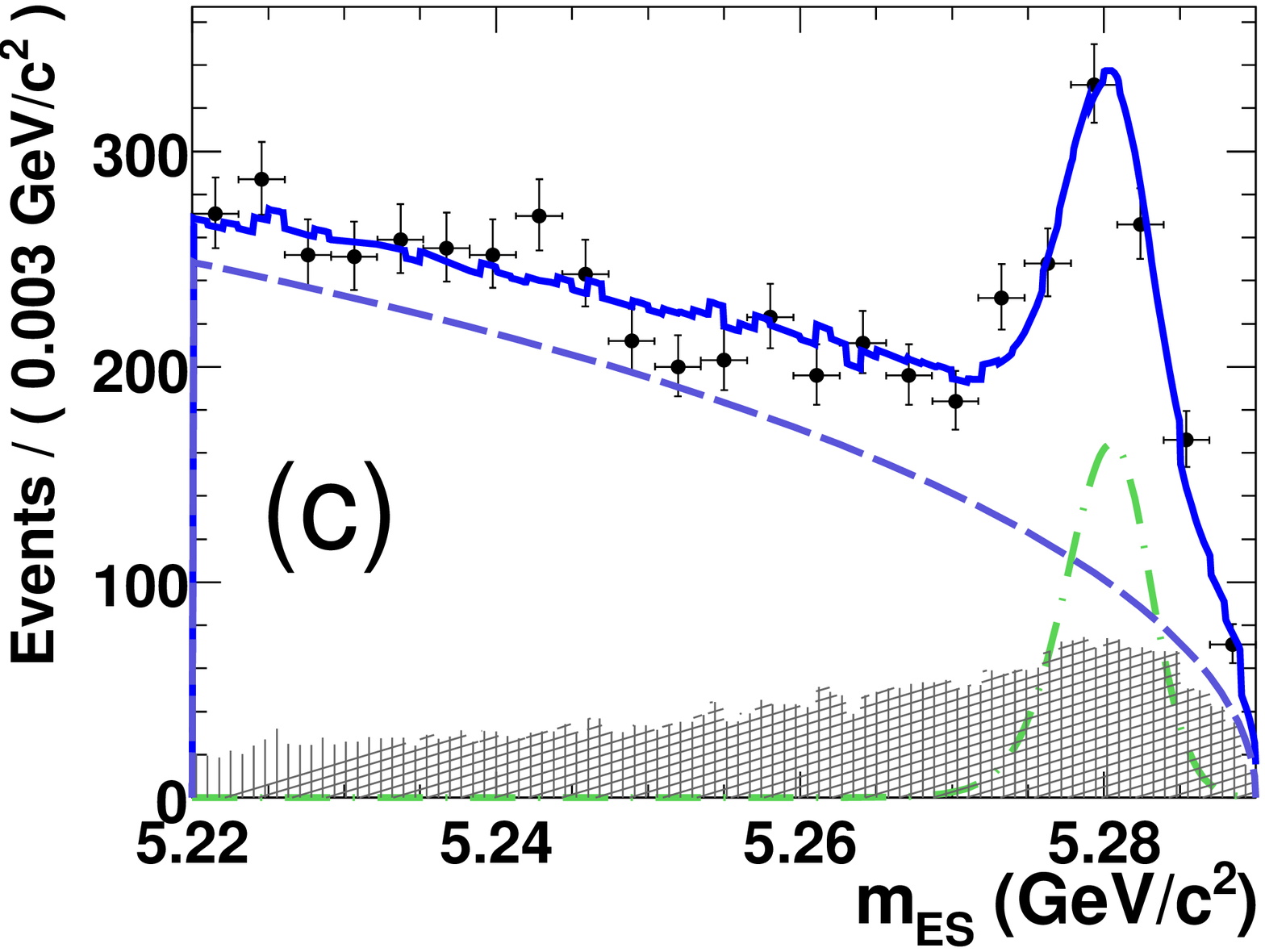}
  \includegraphics[width=0.46\linewidth,keepaspectratio]{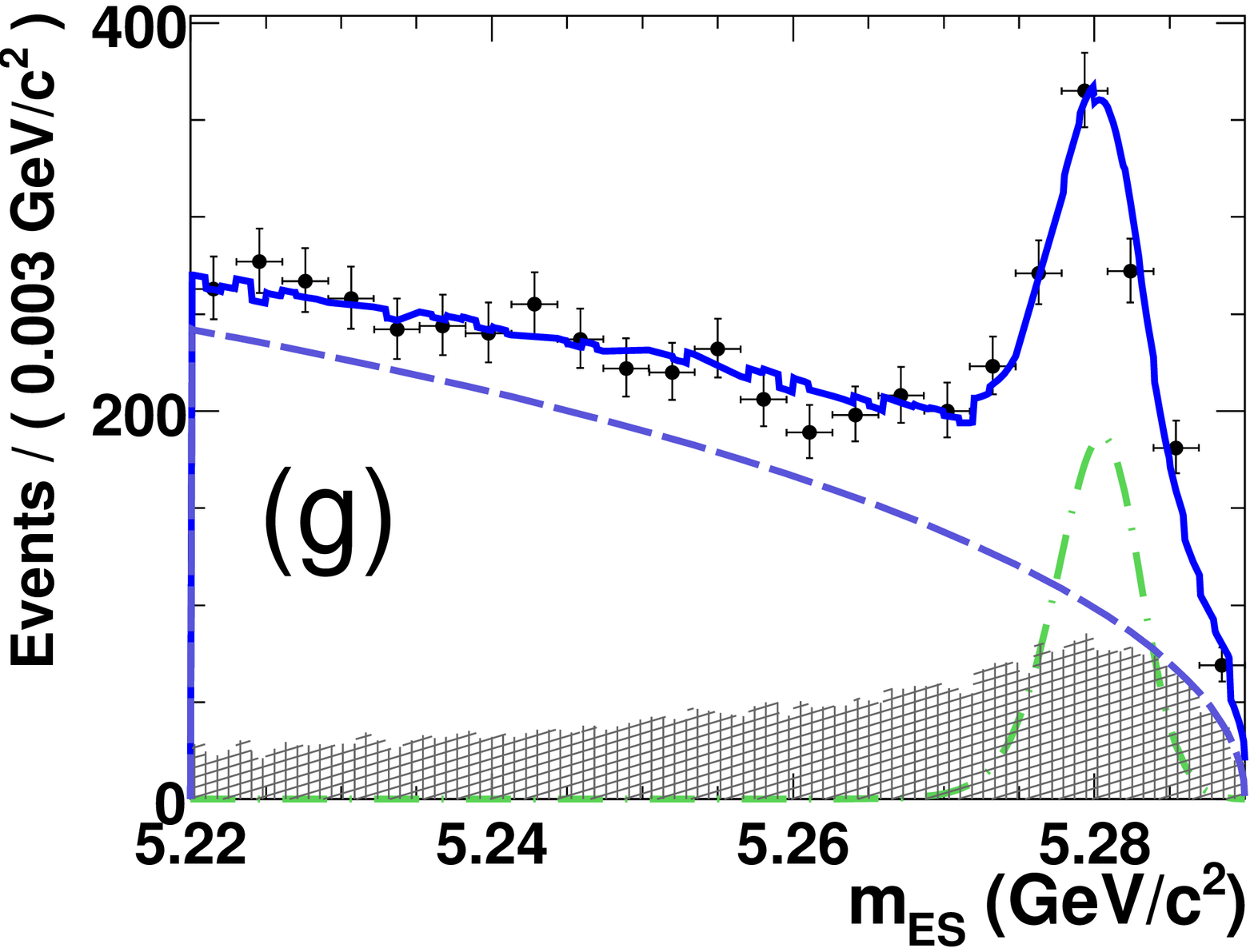}

  \includegraphics[width=0.46\linewidth,keepaspectratio]{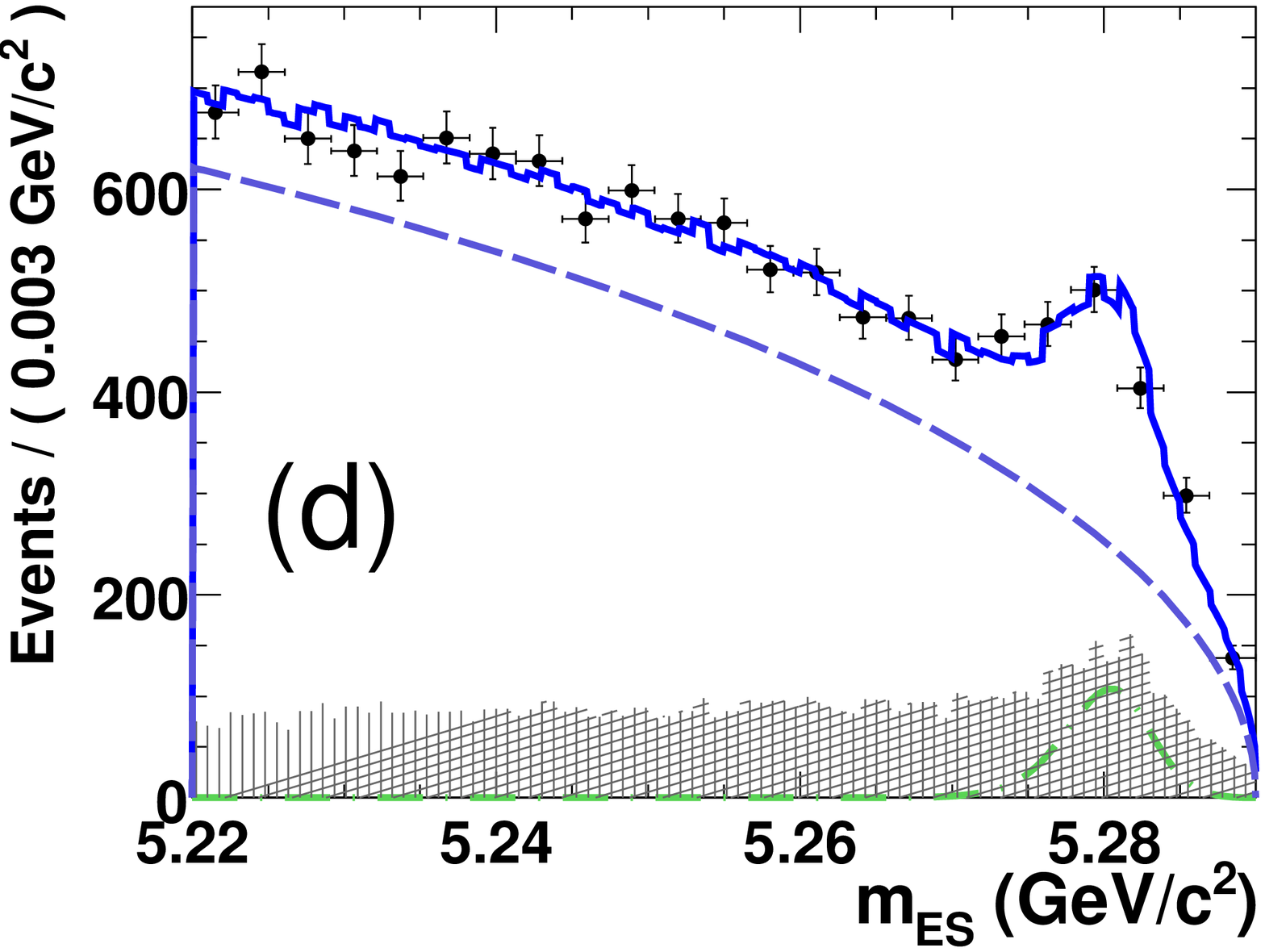}
  \includegraphics[width=0.46\linewidth,keepaspectratio]{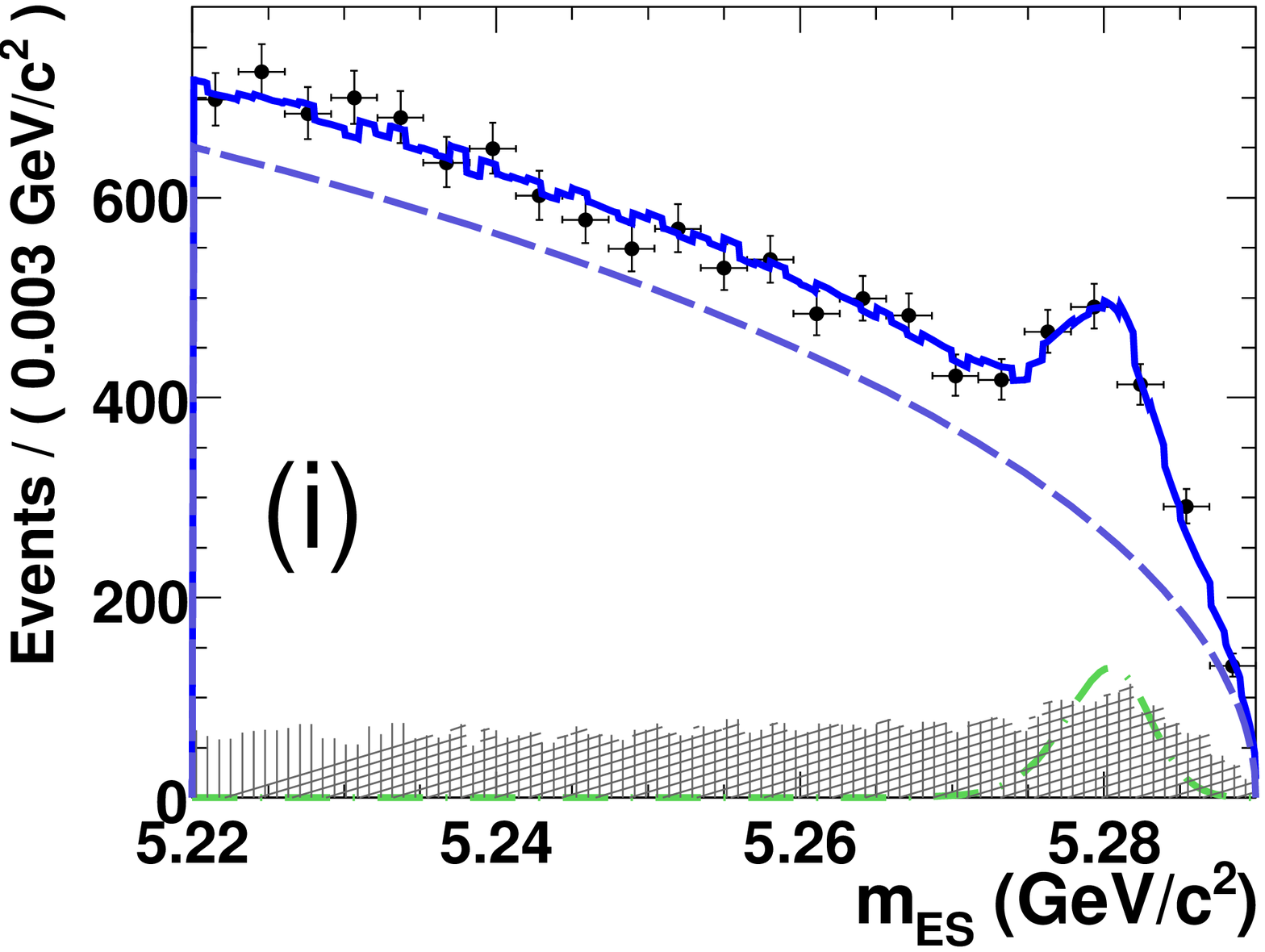}

  \caption{\label{fig:Proj_FlavorFourRange}Fits to the \mes\ distribution in data for \btosgamma\ events in
  $M_{\Xs}$ region
  (a) [0.6, 1.1], (b) [1.1, 1.5], (c) [1.5, 2.0], (d) [2.0, 2.8],
  and \bbartosbargamma\ events in $M_{\Xs}$ region
  (e) [0.6, 1.1], (f) [1.1, 1.5], (g) [1.5, 2.0], (h) [2.0, 2.8],
 The dashed line shows the shape of the continuum,
 dotted-dashed line shows the fitted signal shape
 and the dotted line shows the \BB\ and cross-feed shape.
}
\end{center}
\end{figure}
The background surviving the final selection can be
attributed to one of three sources: continuum events,
\BB\ events other than \btosgamma\ decays 
(referred to as generic \BB ),
and ``cross-feed events'', defined as events containing a
\btosgamma\ decay, but in which the true 
decay
was not correctly reconstructed. The shape of the cross-feed
and \BB\ background is described by
a binned PDF, determined from MC with 1 \mevcc\ binning.

The continuum background is described by an ARGUS function
\cite{Albrecht:1986nr} determined from a fit to the off-resonance data.
In this fit, the \mes\ distribution is shifted to have the
same end-point as that of the on-resonance data.
%
%

\begin{figure}[t]
\begin{center}
  \includegraphics[width=0.46\linewidth,keepaspectratio]{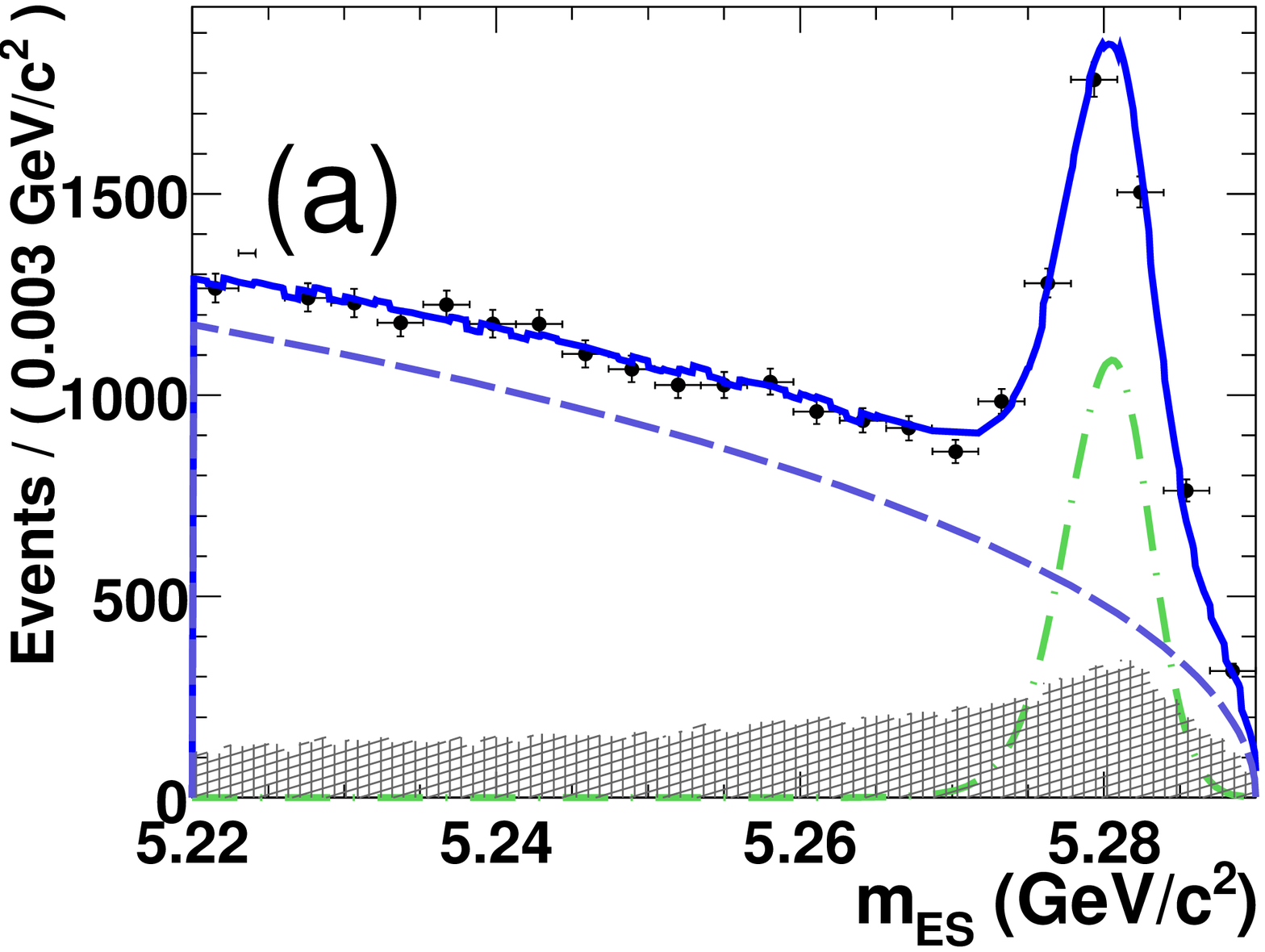}
  \includegraphics[width=0.46\linewidth,keepaspectratio]{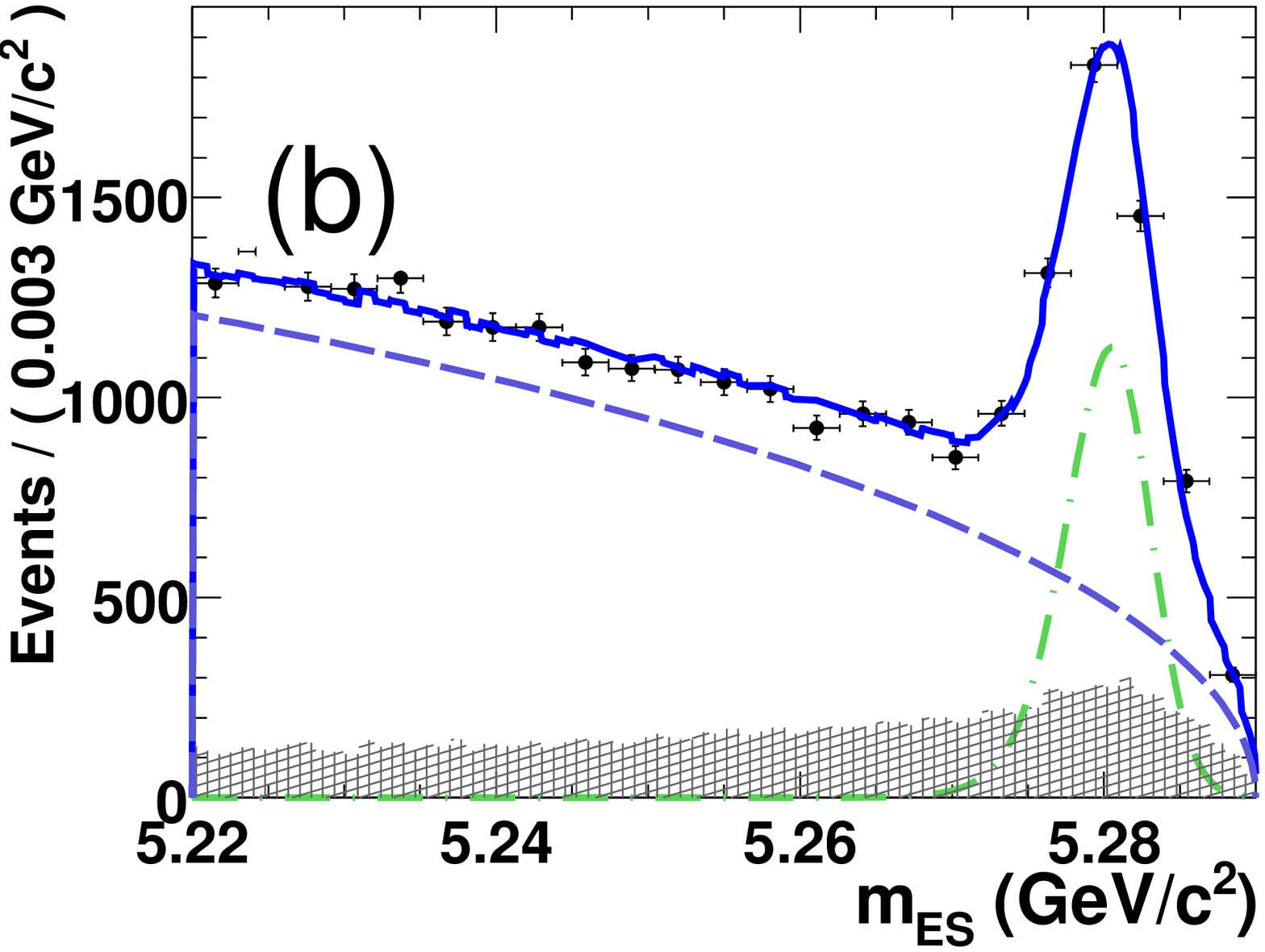}
  \caption{\label{fig:Proj_FlavorFullRange}Fits to the \mes\ distribution in data for (a) \btosgamma\ events 
  and (b) \bbartosbargamma\ events in the entire  $M_{\Xs}$ region.
 The dashed line shows the shape of the continuum, the 
 dotted-dashed line shows the fitted signal shape, 
 and the dotted line shows the \BB\ and cross-feed shape.
}
\end{center}
\end{figure}

In the maximum-likelihood fit, all parameters are fixed
with the exception of the normalizations of the various components 
as well as $\mu_0$, which is determined from fitting
the data, since the peak position is not well modeled in the
MC simulation. The signal, \BB\ and cross-feed shapes are constrained
by the MC, while the continuum background shape is fixed to that of 
off-resonance data.
The shapes of the distributions are assumed to be the same for \B\ and
$\Bbar$ candidates, with the exception of the \BB\ and cross-feed
background, which are allowed to vary between \b\ and \bbar\
in order to eliminate the possibility of a false \CP\ asymmetry.
In Figure \ref{fig:Proj_FlavorFourRange}
we present the final fits to the \mes\ distributions
for \btosgamma\ and \bbartosbargamma\ events for
the four $M_{\Xs}$ sub-regions.
As expected, the signal to background ratio decreases 
from lower to higher $M_{\Xs}$ regions.
In Figure \ref{fig:Proj_FlavorFullRange},
we present the final fits to the \mes\ distribution for \btosgamma\ and
\bbartosbargamma\ events for the entire $M_{\Xs}$ region.

\begin{table*}
\centering
\begin{tabular}{|l|c|c|c|c|c|c|}
\hline
\hline
\noalign{\vskip1pt}
$M_{Xs}$ & \multirow{2}{*}{$\frac{N_b-N_{\bbar}}{N_b+N_{\bbar}}$} & \multirow{2}{*}{$A_{det}$} &  \BB\ and cross-feed & Continuum & \multirow{2}{*}{$A_{CP}$}  \\
(\gevcc) & & & model syst & model syst     &        \\
\hline
\hline
0.6--1.1 &$\phantom{-}0.015 \pm 0.029 $&$ \phantom{-}0.005\pm0.014$ &  0.002  & 0.004  & $\phantom{-}0.010\pm0.029\pm0.015$ \\
1.1--1.5 &$-0.003 \pm 0.049           $&$-0.003\pm0.015$            &  0.003  & 0.004  & $\phantom{-}0.000\pm 0.049\pm0.016$\\
1.5--2.0 &$-0.064 \pm 0.077           $&$-0.017\pm0.010$            &  0.010  & 0.002  & $-0.047\pm 0.077\pm0.014$ \\
2.0--2.8 &$-0.097 \pm 0.180           $&$-0.002\pm0.005$            &  0.070  & 0.168  & $-0.077\pm 0.180\pm0.182$\\
\hline
0.6--2.8 &$-0.018 \pm 0.030           $&$-0.007\pm0.005$           &  0.012 & 0.006    & $-0.011\pm 0.030\pm0.014$ \\
\hline
\hline
\end{tabular}
\caption{\label{tab:SignalFractions}
For each $M_{\Xs}$ bin, we present the fitted \CP\ asymmetry:
$(N_b-N_{\bbar})/(N_b+N_{\bbar})$, the flavor-bias
of the detector: $A_{det}$, the systematic error arising from
the \BB\ and cross-feed modeling and the systematic
error arising from the continuum background modeling. The last
column shows the final results for the \CP\ asymmetries.
}
\end{table*}

The direct \CP\ asymmetry is calculated as
\begin{equation}
\label{equa:ACP}
A_{CP}=\frac{1}{\langle D\rangle} \left( \frac{N_b-N_{\bbar}}{N_b+N_{\bbar}}-\Delta D \right) -A_{det}
\end{equation}
where $N_b$ and $N_{\bbar}$ are the yields of the \btosgamma\
and \bbartosbargamma\ signals respectively. 
$A_{det}$, described in details below,
is the flavor bias caused by the detector responses to positively
and negatively charged particles.
Table \ref{tab:SignalFractions} presents the fitted values for
$(N_b-N_{\bbar})/(N_b+N_{\bbar})$.

$\Delta D = (\bar{\omega}-\omega)$
is the difference in the wrong-flavor
fraction between \b\ and \bbar\ decays, and
$\langle D\rangle=1-(\bar{\omega}+\omega)$
is the dilution factor from the average wrong-flavor fraction.
The small wrong-flavor fraction $\bar{\omega}$ ($\omega$), defined to
be the fraction of \bbar (\b) reconstructed as the opposite flavor,
is due to charged pions misidentified as charged kaons.
Using the particle
misidentification rate measured in control samples in data 
we calculate  $\Delta D =(5\pm4)\times10^{-5}$
and $1-\langle D\rangle = (5.4\pm0.1)\times10^{-3}$.

The flavor bias of the detector $A_{det}$ is due to asymmetric
$K^+$, $K^-$ interaction cross-sections in the detector
at low momenta. Such an asymmetry could
produce a false \CP\ asymmetry in the signal events.
We perform a measurement of $A_{det}$ in data using two independent
methods. The first approach determines  this asymmetry from events
in the \mes\ sideband, $5.22<\mes<5.27 \gevcc$, which is dominated
by continuum background with no expected \CP\ asymmetry.
The second approach uses a control sample where
we replace the high-energy photon from the \B\ decay with a high-energy
\piz  with $p_{CM}\ge$ 1.6\gevc. 
The same selection criteria used 
in the signal selection are applied, except for 
\piz\ and $\eta$ veto requirements, and  the \CP\ asymmetry in the \mes\
sideband is measured. In both control samples 
we apply appropriate weights
to the events to ensure that the fraction of each reconstructed
final state is identical to that in the signal sample.
We find the \CP\ asymmetry measured using
both of these approaches to be nearly identical, 
and average the two measurements 
to obtain $A_{det}=-0.007 \pm 0.005$. The mean value is used to shift the 
$(N_b-N_{\bbar})/(N_b+N_{\bbar})$ mean value, while the error contributes to 
 the systematics. 
 The values of  $A_{det}$ computed in each \Xs\ mass region  are reported in Table \ref{tab:SignalFractions}.

The shape of the \BB\ and cross-feed background, determined from MC,
is also a potential source of flavor bias in the fit to the data.
This background peaks broadly in the signal region, and a small
shape difference as a function of flavor could create a false \CP\
asymmetry in the signal. 
We measure the size of this effect by correcting the \BB\ and cross-feed
shapes separately. 
The high-energy \piz\ control
sample  is used to study the uncertainty of the
\BB\ background shape. We use the differences found between the
data and MC \mes\ shapes in this control sample to correct the
nominal \BB\ background shape built from the MC.
The biggest uncertainty in the cross-feed shape is due to the fact that
JETSET does not reproduce the observed fragmentation structure of data.
We thus correct the simulation shape using the fragmentation previously determined
from \babar\ data~\cite{Aubert:2005cua}.
We then construct new \b\ and \bbar\ binned PDFs using these corrected
cross-feed and \BB\ events and fit the data a second time with them.
The difference between the nominal \ACP\ and \ACP\ from this fit,
shown in Table \ref{tab:SignalFractions}, is used as the systematic
error from shape modeling of the \B\ background.

The systematic error arising from the continuum background modeling
is determined by varying the ARGUS shape parameters within the
experimental errors, and  is found to be 0.006 for the combined $M_{\Xs}$ region.
Systematic errors due to possible differences in the  
signal shape  between \b\ and \bbar\ events, \CP\ content of the peaking background,  
and possible contaminations from $b\to d \gamma$ decays 
are all found to be negligible.  Contributions from $\langle D\rangle$,
$\Delta D$ and signal modeling are neglected due to their small impact on \ACP.
The dominant systematic errors  are therefore due to the uncertainties
in the flavor bias of the detector and the background shapes as described
above. 

The total systematic errors are calculated as the sum in
quadrature of errors on $A_{det}$, systematic errors arising from the continuum,
\BB\ and cross-feed shape modeling. 
The results are shown in Table \ref{tab:SignalFractions}.

In summary, we measure the direct \CP\ asymmetry in \btosgamma\ to be
\ACP = $-0.011\pm 0.030\pm0.014$ in the region $0.6<M_{\Xs}<2.8 \gevcc$. 
This result represents the most accurate measurement of this quantity to date.
The measurement is consistent with zero \CP\ asymmetry and with 
the SM prediction. 
The \CP\ asymmetry in each $M_{\Xs}$ region considered in our study is also consistent with zero.

We are grateful for the excellent luminosity and machine conditions
provided by our \pep2\ colleagues,
and for the substantial dedicated effort from
the computing organizations that support \babar.
The collaborating institutions wish to thank
SLAC for its support and kind hospitality.
This work is supported by
DOE and NSF (USA),
NSERC (Canada),
CEA and
CNRS-IN2P3
(France),
BMBF and DFG
(Germany),
INFN (Italy),
FOM (The Netherlands),
NFR (Norway),
MES (Russia),
MEC (Spain), and
STFC (United Kingdom).
Individuals have received support from the
Marie Curie EIF (European Union) and
the A.~P.~Sloan Foundation.

\bibliography{bsgamma} 

\end{document}